# Polar state of $0.67BiFeO_3$–$0.33BaTiO_3$ near the morphotropic phase boundary

Yongxing Wei,[1,a] Jiahao Shen,[1] Chenxing Bai,[1] Changqing Jin,[1] Weitong Zhu,[2] Ye Tian,[3] Gang Xu[1]

**AFFILIATIONS**

[1] Shaanxi Key Laboratory of Optoelectronic Functional Materials and Devices, School of Materials and Chemical Engineering, Xi'an Technological University, Xi'an 710021, China

[2] Electronic Materials Research Lab, Key Laboratoryof Ministry of Education and International Center for Dielectric Research, School of Electronic and InformationEngineering, Xi'an Jiaotong University, Xi'an 710049, China

[3] School of Materials Science and Engineering, Shaanxi University of Science & Technology, Xi'an 710021, China

**ABSTRACT**

The symmetric studies on the structure-property relationship of the unpoled and poled states of $0.67BiFeO_3$–$0.33BaTiO_3$ (0.67BF–0.33BT) were conducted to understand the origin of the morphotropic phase boundary (MPB) in BF–BT. A typical relaxor-type dielectric anomaly was observed ($T_f$, ~627 K). The remnant polarization ($P_r$) and maximum value of electro-strain ($S_m$) increase clearly during heating ($P_r$, ~40 μC/cm$^2$; Sm, 0.191 % under 40 kV/cm at 453 K). The first-cycle electro-strain loops indicate the difference in the polar state between 0.67BF–0.33BT and $0.94BiNaTiO_3$–$0.06BaTiO_3$. Both the unpoled and poled samples have the similar frequency dispersion behaviors. Even in the poled samples, the transition between the ergodic relaxor state and ferroelectric-like state does not involve a clear dielectric anomaly. Analyses based on the Rietveld refinement of XRD patterns, bright-field images and selected-area electron diffractions (SAED) demonstrated that the formation of the long-range ferroelectric domains was difficult under the poling field.

---
[a]Author to whom correspondence should be addressed:
weiyx1985@gmail.com

# I . INTRODUCTION

BiFeO$_3$–BaTiO$_3$ (BF–BT), first reported by Ismailzade et al. in 1981,[1] aroused great interest because of its multiferroic properties in the initial stage.[2–4] In 2009, the good piezoelectric coefficient ($d_{33}$, 116 pC/N) and high Curie temperature ($T_c$, > 873 K) of 0.75BF–BT were obtained by the Mn modification.[5] The studies by Wei et al. and Yang et al. confirmed that the optimum piezoelectric properties were realized in compositions (BaTiO$_3$ content, ~30 mol%) near the morphotropic phase boundary (MPB).[6–8] In 2015, the high piezoelectric performance ($d_{33}$, 402 pC/N) and large electrostrain ($S_m$, ~0.2 % under 50 kV/cm) were reported in the water quenched BF–BT ceramics.[9] Advanced with the high depolarization temperature ($T_d$, > 673 K) and good thermal stability, BF–BT has become one of the most promising lead-free options .[10–19]

Despite the intensive studies, the controversy on the origin of MPB has continued. The early study suggested a cubic phase appeared when the BaTiO$_3$ content reached 33 mol%.[1,2] Leontsev et al. held that the BaTiO$_3$-rich phase should be pesudocubic due to the presence of the ferroelectric and piezoelectric properties.[3] The study by Lee et al. highlighted the similarity in MPB between BF–BT and Pb(Zr,Ti)O$_3$.[9] In 2017, Wang et al. found that the electrostrain of 0.7BF–0.3BT increased when heated.[13] A field induced relaxor-ferroelectric transition was presumed to interpret this enhancement. However, the in-situ synchrotron X–ray diffraction (XRD) in combination with Rietveld refinement of pseudocubic BF–BT suggested that there was no clear peak splitting when the poling field was imposed.[20,21]

Is the initial state of 0.67BF–0.33BT (a) a nonergodic relaxor state that develops to a long-range ferroelectric state under the electric field, (b) a ferroelectric state with small tetragonal distortion, (c) or a ferroelectric state with nanodomains adopted pseudocubic symmetry on the global length scale (Fig. 1)? The structure changes based on the above three models are too similar to be resolved by X-ray diffractions. The analyses of the first-cycle electrostrain loop and structure-property relationship of the poled sample are important. If the transition from the nonergodic relaxor state to the ferroelectric state occurs, we could observe its effect on the first-cycle electrostrain loop, domain structure and dielectric properties like $Bi_{0.5}Na_{0.5}TiO_3$-based ceramics.[22–25] In this study, the initial electrostrain loops of 0.67BF–0.33BT were obtained and poling effects on the structural and electrical properties were analysed to understand its polar state.

## II. EXPERIMENTAL PROCEDURE

Solid solution ceramics of 0.67BF–0.33BT were prepared by the solid-state reaction method, with the starting reagents of $Bi_2O_3$, $BaCO_3$, $Fe_2O_3$ and $TiO_2$. 1 mol% $MnO_2$ was added in order to increase the insulation. The sintering temperature was 1000 ºC. For the electric measurements, the silver paste was coated and fired at 550°C for 10 minutes. The dielectric data were measured using an LCR meter (4294 A, Agilent, Santa Clara, America) with a temperature-controlled cell. The ferroelectric properties and the electrostrain responses at 1 Hz were measured using a ferroelectric test system (TF Analyzer 2000E, aixACCT, Aachen, Germany). The ceramics were poled in silicone oil for 10 minutes under a dc field of 40 kV/cm at 298 K for the piezoelectric measurement. The piezoelectric coefficients ($d_{33}$) were measured using a piezo-$d_{33}$ meter (ZJ–3AN; Institute of Acoustics, Beijing, China). The crystal structures were detected using an X–ray diffractometer (XRD, D2 PHASER, Bruker AXS, Karlsruhe, Germany). The Rietveld refinement analysis (using the FullProf software) was performed to study the poling effect on the structure. The bright-field images and selected-area electron diffraction SAED patterns were obtained by the transmission electron microscopy (TEM, JEM-2100Plus; JEOL, Okyo Metropolis, Japan).

## III. RESULTS AND DISCUSSION

Fig. 2(a) shows the relative permittivity ($\varepsilon_r$) and dielectric loss of 0.67BF–0.33BT as a function of temperature at various frequencies. A broad and frequency dependence dielectric anomaly at approximately 700 K reveals a strong relaxor behavior in a high-temperature range. The relationship between the temperature ($T_m$) for the maximum value of $\varepsilon_r$ and measuring frequency can be well described by the Vogel-Fulcher law (Fig. 2b), giving a freezing temperature ($T_f$) of ~627 K, ~50 K higher than that reported by Zheng et al.[11,26] In relaxor ferroelectrics, the polar nano regions (PNRs) appear at the Burns temperature ($T_B$) where the relationship between the reciprocal permittivity and temperature departs from the Curie-Weiss law when cooled.[11] The high temperature data at 100 kHz were used to define the $T_B$ value (Fig. S1). The $T_B$ value was found to be ~820 K, ~95 K higher than $T_m$ (725 K at 100 kHz).

The relationship between the piezoelectric coefficient ($d_{33}$) and annealing temperature indicates that the depolarization temperature is around 650 K, close to $T_f$ (Fig. 2c). The polarization and electro-strain responses during heating were studied to better understand the ferroelectric-like state of 0.67BF–0.33BT (Fig. 2d–e). The detailed experimental results are shown in Fig. S2 in the supporting information. The remnant polarization ($P_r$) increases clearly with the temperature, accompanied by a reduction in the coercive field ($E_c$). Above 398 K, the conduction contributes much to the polarization response. We obtained the intrinsic ferroelectric hysteresis loop by deducting the conduction effect, assuming that the relation of leakage current density and electric field is linear.[27] The intrinsic $P_r$ value of 0.67BF–0.33BT is ~40 μC/cm² at 453 K, much higher than that of $Bi_{0.5}Na_{0.5}TiO_3$.[24] Most importantly, the heating also induces an enhancement of the electrostrain. At 453 K, the maximum value of unipolar strain ($S_m$) and large-signal piezoelectric coefficient ($d_{33}^*$) are ~0.191 % and ~477 pm/V, respectively. The trends of the polarization and electrostrain for 0.67BF–0.33BT when heated are identical with those reported in 0.7BF–0.3BT.[13,28]

The first-cycle ferroelectric hysteresis loop and electrostrain loop of 0.67BF–0.33BT at room temperature are shown in Fig.3a. The strain changes slowly when the poling field ($E$) is lower than the coercive field ($E_c$). However, when the poling field approaches $E_c$, the polarization and strain raise quickly. A large remnant strain could be observed when the poling field was removed. The shape of the first-cycle ferroelectric hysteresis loop and electro-strain loop of 0.67BF–0.33BT is similar to that observed in normal ferroelectrics.[29] At 453 K, the strain increases clearly when the electric field is imposed, which is different from that at room temperature. The initial curve of strain and electric field is found to obey the following expression

$$S = aE + bE^2,$$

where $S$ is the strain, $E$ is the electric field, and $a$ (0.0036 ± 0.0001) and $b$ (0.000110 ± 0.000006) are fitting parameters. The finding suggests the initial state of

0.67BF–0.33BT differs from typical nonergodic relaxors.[30]

The poling effect on the relationship between the dielectric behavior and frequency is plotted in Fig.4a. The low dielectric loss in a low-frequency range suggests that the leakage current is effectively suppressed near the room temperature by Mn-modification. The relative permittivity of the unpoled sample is frequency dependent. In order to estimate the frequency dispersion, the slope of relative permittivity versus log frequency plots was calculated. The slope of the unpoled sample is approximately –21. After poling, the relative permittivity increases clearly but the frequency dispersion behaviors are nearly unchanged (slope, approximately –23). The temperature dependence of relative permittivity suggests that the $T_m$ value shifts to the high-temperature side after poling (Fig. 4b). At 100 kHz, $T_m$ shifts from 725 K to 729 K after poling. The shift of $T_m$ induced by poling was also observed in $PbMg_{1/3}Nb_{2/3}O$–$PbTiO_3$.[31] Most importantly, there is no clear dielectric anomaly related to the transition point between ferroelectric-like state and relaxor state in the poled ceramic.

The XRD patterns of unpoled and poled powdered samples for 0.67BF–0.33BT are shown in Fig. 5(a) and (b). A cubic-like perovskite phase was found both in unpoled and poled ceramics, consistent with the in-situ synchrotron radiation XRD analyses.[20,21] The reflection peak shifts to the low-angle side after poling. Rietveld refine analysis suggests that the structures could be well described by the cubic symmetry with the space group of $Pm\bar{3}m$. The poling leads to a 0.14 % increase in the lattice parameter (*a*). The increase in the lattice parameter by dc electric field poling is corresponding to the initial electrostrain loop (Fig. 3a).

The bright-field TEM images of many grains were examined to understand the micro polar order. No viable ferroelectric domains were detected both in the unpoled (Fig. 5c) and poled state (Fig. 5d). It reveals the difficulty in the formation of long-range ferroelectric domains under the poling field.

To further understand the poling effect on the structure, the SAED patterns with [110] zone axis were obtained (Fig. 5e and f). The absence of the super-lattice reflection spots $1/2(111)_c$ suggests that there is no ordered rotation of the octahedral both in the initial and poled state of 0.67BF–0.33BT.[31] That is, the R3c structure should be eliminated if the rombohedral distortions are present on the local scale. The increase in the tolerance factor ($t = 0.968$ for 0.75BF–0.25BT, $t = 0.970$ for $Bi_{1/2}Na_{1/2}TiO_3$ and $t = 0.978$ for 0.67BF–0.33BT) and the degree of disorder with increasing $BaTiO_3$ content leads to the difficulty in the ordered rotation of the octahedral.

Benefitting from the high insulation by Mn-modification, the intrinsic electrical properties were obtained. we compared 0.67BF–0.33BT with $0.64PbMg_{1/3}Nb_{2/3}O_3$–$0.36PbTiO_3$ (0.64PMN–0.36PT)[32] and $0.94Bi_{1/2}Na_{1/2}TiO_3$–$0.06PbTiO_3$ (0.94BNT–0.06BT)[22,23,33–35] near MPB (Table 1) to better understand its structural and electrical characteristics. The eight main differences in 0.67BF-0.33BT are summarized as below.

(1) Both the structures of the unpoled and poled states could be well understood as the pseudocubic symmetry.[20,21]

(2) No visible ferroelectric domains could be detected in the unpoled state. The poling field hardly triggers the formation of the long-range ferroelectric domains.

(3) There are no clear differences in the frequency dispersion behaviors between

the unpoled and poled states.

(4) The frequency dependence of $T_m$ is obvious and well fitted by the V-F functions.

(5) The difference of $T_m$ and $T_B$ is about ~95 K, higher than that in 0.64PMN–0.36PT but lower than that in 0.94BNT–0.06BT.

(6) The relationship between the relative permittivity and temperature in the poled sample lacks the clear anomaly relates to the transition from the relaxor state to the ferroelectric-like state.

(7) Despite the large remnant strain, the initial strain loops differ from those in typical nonergodic ralaxors, without the evidence of the electric-field-induced change from the nonergodic relaxor ferroelectric state to long-range ferroelectric state.

(8) The heating leads to the increase in the electrostrain and the polarization, even the maximum measuring temperature is ~200 K lower than $T_d$.

Our studies reveal the complexity of 0.67BF-0.33BT. This is possibly caused by the mixtures of the ferroelectrically active and nonferroelectrically active cations both on the A site and B site. The origin of the polar state remains unclear and should be further studied.

## IV. CONCLUSIONS

In summary, the first-cycle electrostrain loops and poling effects on the dielectric properties, structures and domain structures of 0.67BF–0.33BT were studied. The initial strain loops are different from that of 0.94BNT-0.06BT. In addition, the poling does not lead to clear changes in the frequency dispersion behavior and crystal symmetry. The bright-field image of the poling sample suggests that the formation of the long-range domains is difficult when the poling field is imposed. There are clear differences in MPB among BF–BT, BNT–BT and PMN-PT. This study can improve the understanding of the polar state in BF-BT near MPB.

**SUPPLEMENTARY MATERIAL**

See supplementary material for the Curie–Weiss fit (Fig. S1) and polarization, bipolar and unipolar responses at various temperatures (Fig. S2) of 0.67BF–0.33BT.

**ACKNOWLEDGEMENTS**

This work was financially supported by the National Natural Science Foundation of China (Project No. 11704301), the Natural Science Basic Research Plan in Shaanxi Province of China (Program No. 2018JQ1092), the Shaanxi Provincial Education Department Program (Program No.19JK0398) and the President's Fund of Xi'an Technological University (Project no. XAGDXJJ18006).

**DATA SHARING POLICY**

The data that support the findings of this study are available from the corresponding author upon reasonable request.

**TABLE CAPTION**

**TABLE 1**. Structural and electrical characteristics of 0.64PMN–0.36PT, 0.94BNT–0.06BT and 0.67BF–0.33BT. FE, ferroelectric.

**FIGURE CAPTIONS**

**FIGURE 1**. Schematic of the three different phase diagrams in BiFeO$_3$–BaTiO$_3$ (BF–BT). $T_{F-R}$, $T_f$ and $T_B$ represent the transition point between the nonergodic relaxor state and ferroelectric state, freezing temperature and Burns temperature at which the polar nanoregions PNRs appear, respectively.

**FIGURE 2**. (a) Temperature dependence of relative permittivity $\varepsilon_r$ and dielectric loss $tan\delta$ at selected frequencies, (b) Vogel-Fulcher law fitting, (c) relation of piezoelectric coefficient and annealing temperature and (d) polarization, (e) bipolar and (f) unipolar responses at various temperatures of 0.67BF–0.33BT.

**FIGURE 3**. First-cycle ferroelectric hysteresis loops and electro-strain loops at (a) 298 K and (b) 453 K, inset of Fig. 3 (b) shows the fitting for the initial curve of the strain and electric field.

**FIGURE 4**. Poling effect on the (a) frequency and (b) temperature dependence of dielectric behaviors

**FIGURE 5**. Rietveld fitted powder XRD patterns of (a) unpoled and (b) poled state, the black cycles represent the observed pattern, the red continuous line is correspond to the fitted pattern, the blue vertical bars point the Bragg peak positions, the magenta continuous line at the bottom represents the difference between the observed and fitted pattern. Poling effects on the (c), (d) bright-field images and (e), (f) SAED patterns.

**TABLE 1**. Structural and electrical characteristics of 0.64PMN–0.36PT, 0.94BNT–0.06BT and 0.67BF–0.33BT. FE, ferroelectric.

| Composition | 0.64PMN-0.36PT[32] | 0.94BNT-0.06BT[22,23,33–35] | | 0.67BF-0.33BT | |
|---|---|---|---|---|---|
| | | unpoled | poled | unpoled | poled |
| Structure | Tetragonal | Pseudocubic | Tetragonal | Pseudocubic | Pseudocubic |
| Domains structures | Long-range FE domains | PNRs | Long-range FE domains | No long-range FE domains | No long-range FE domains |
| Slope of $\varepsilon_r$ vs. ln($f$) | — | −56 | −11 | −21 | −23 |
| Frequency dependence of $T_m$ | Weak | Weak | | Strong | |
| Difference between $T_m$ and $T_B$ | ~30 K | ~240 K | | ~95 K | |
| Initial strain loop | — | Evidence for a phase transition | | No evidence for a phase transition | |

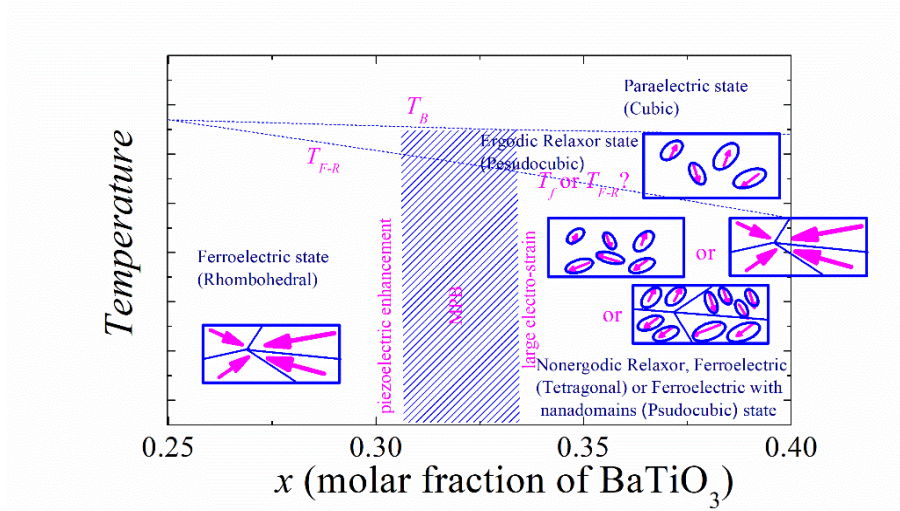

**FIGURE 1**. Schematic of the three different phase diagrams in $BiFeO_3$–$BaTiO_3$ (BF–BT). $T_{F-R}$, $T_f$ and $T_B$ represent the transition point between the nonergodic relaxor state and ferroelectric state, freezing temperature and Burns temperature at which the polar nanoregions PNRs appear, respectively.

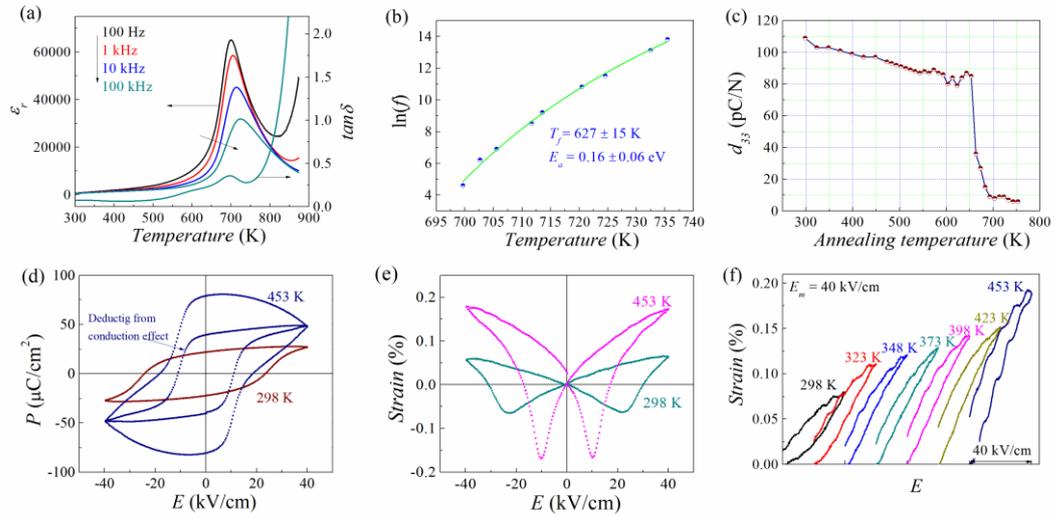

**FIGURE 2**. (a) Temperature dependence of relative permittivity $\varepsilon_r$ and dielectric loss $tan\delta$ at selected frequencies, (b) Vogel-Fulcher law fitting, (c) relation of piezoelectric coefficient and annealing temperature and (d) polarization, (e) bipolar and (f) unipolar responses at various temperatures of 0.67BF–0.33BT.

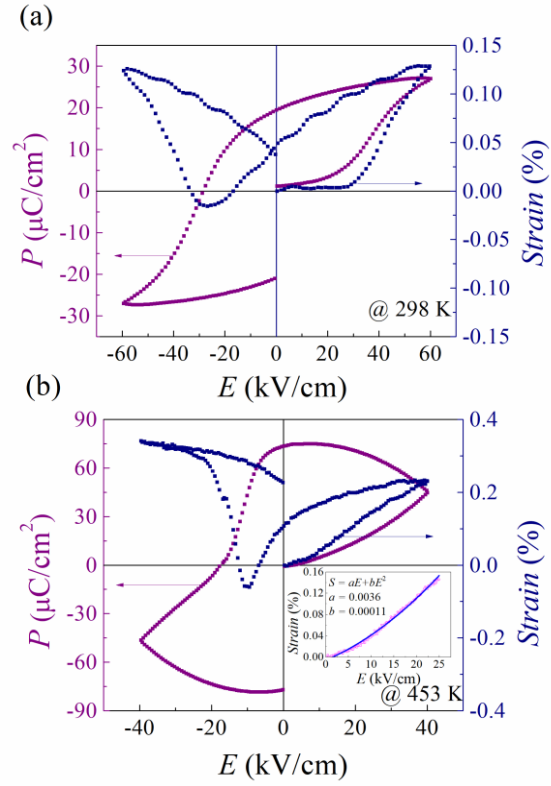

**FIGURE 3**. First-cycle ferroelectric hysteresis loops and electro-strain loops at (a) 298 K and (b) 453 K, inset of Fig. 3 (b) shows the fitting for the initial curve of the strain and electric field.

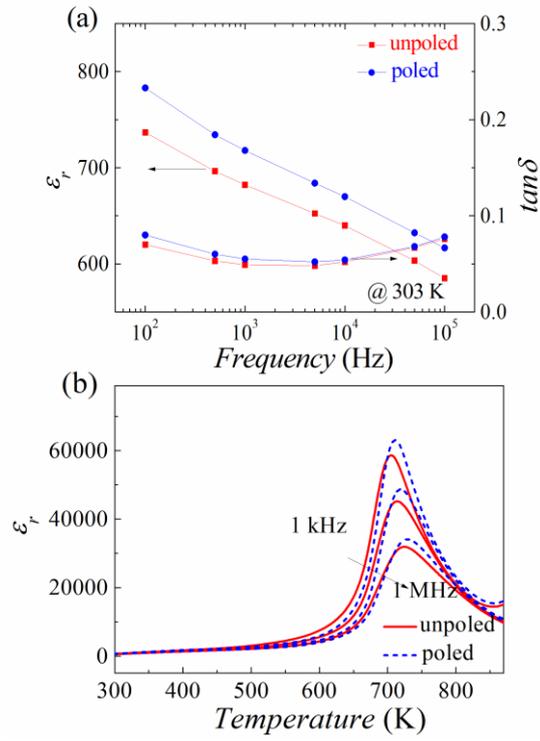

**FIGURE 4.** Poling effect on the (a) frequency and (b) temperature dependence of dielectric behaviors

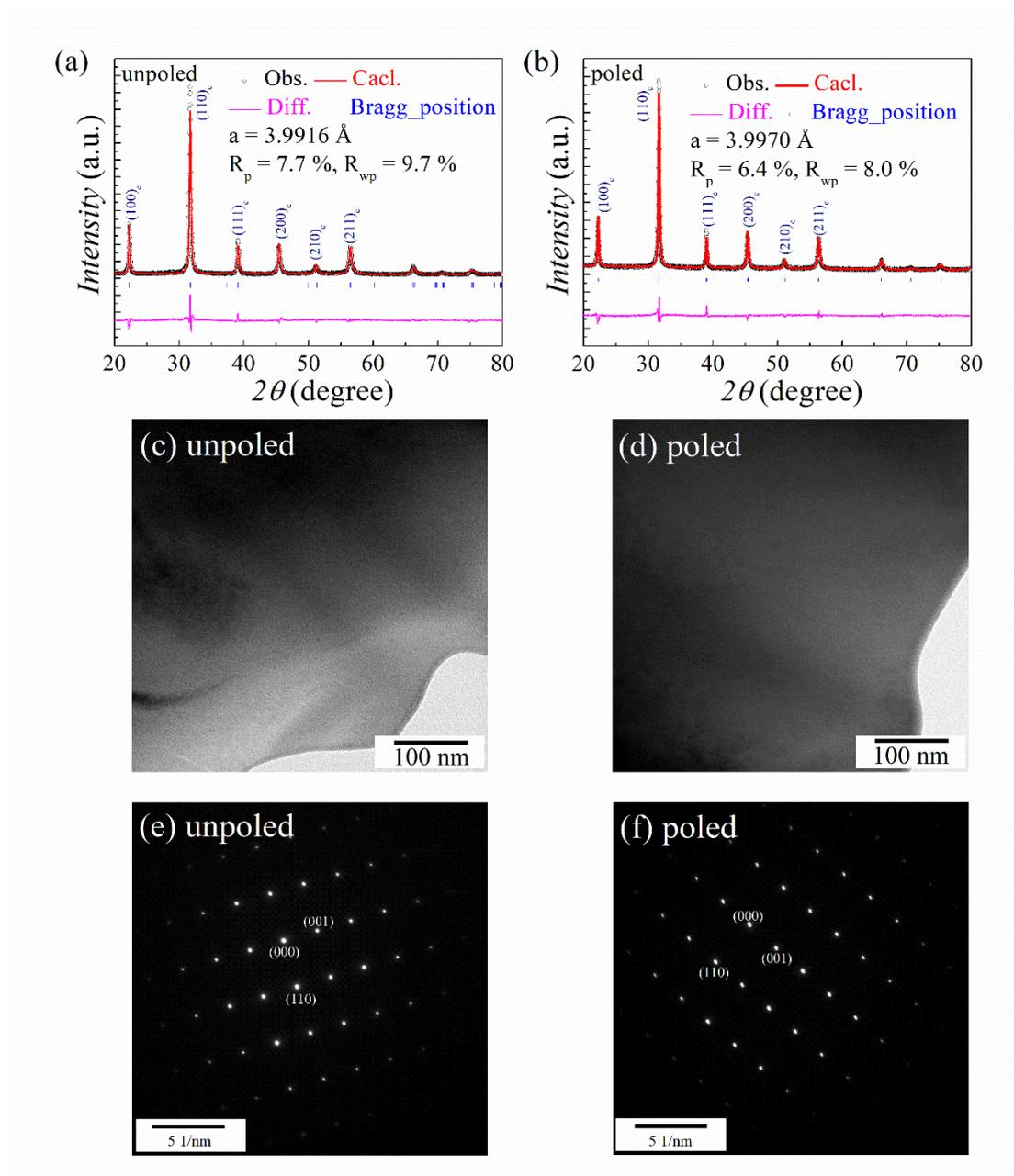

**FIGURE 5**. Rietveld fitted powder XRD patterns of (a) unpoled and (b) poled state, the black cycles represent the observed pattern, the red continuous line is correspond to the fitted pattern, the blue vertical bars point the Bragg peak positions, the magenta continuous line at the bottom represents the difference between the observed and fitted pattern. Poling effects on the (c), (d) bright-field images and (e), (f) SAED patterns.